\documentclass[3p,times,twocolumn]{article}

\usepackage{amssymb}

\usepackage[figuresright]{rotating}
\usepackage{multirow}
\usepackage[T1]{fontenc}
\usepackage[latin1]{inputenc}
\usepackage{pifont}
\newcommand{\cmark}{\ding{51}}

\begin{document}
\begin{titlepage}


\begin{flushright} 
{ Preprint IFJPAN-IV-2021-20 \\  December 2021  
} 
\end{flushright}

\vskip 3 mm
\begin{center}

{ \bf\huge  The $\tau$ lepton Monte Carlo Event Generation --
imprinting  New Physics models with exotic scalar or vector states  \\ into simulation samples}
\end{center}
\vskip 9 mm
\begin{center}
{\bf Sw. Banerjee$^a$, D. Biswas$^a$, T. Przedzinski$^b$, Z. Was$^{\dagger,c}$}
\vskip 3 mm
{\em $^a$ University of Louisville, Louisville, Kentucky, 40292, USA}

{\em $^b$ Institute of Physics, Jagellonian University, 30-348 Krakow, Lojasiewicza 11, Poland }
       
{\em $^c$ Institute of Nuclear Physics, Polish Academy of Sciences, PL-31342 Krakow}

$^{\dagger}$ Speaker
\end{center}
\vskip 6 mm

\begin{abstract}

The Monte Carlo for lepton pair production and $\tau$ decays consist
of {\tt KKMC} for lepton pair production, {\tt tauola} for $\tau$ lepton decays and
{\tt photos} for radiative corrections in decays.

An effort for adaptation of the system for precision data being collected
at the {\tt Belle II} experiment included simulation of  additional light lepton
pairs. Extension to processes where lepton pair is produced through narrow resonances,
like dark photon or dark scalar ($\phi$) resonances, was straight forward.

Modified  programs versions are available in stand-alone format from gitlab repository or through the {\tt basf2} system of
Belle II software. It was explained recently  during the International Workshop on Tau Lepton Physics September, 2021, Bloomington IN. Now we concentrate on
simulations for $\phi$ resonance, a hypothetical object which could be responsible for anomalous moment $g-2$ in $Z-\tau-\tau$ interactions through virtual contributions.   


\vskip 6 mm
\centerline{\bf Presented on STRONG 2020 Virtual Workshop on}
\centerline{\bf Spacelike and Timelike determination of the Hadronic LO contribution to the Muon g-2}
\centerline{\bf November, 2021, Frascati (virtual edition), Italy } 
\end{abstract}

\vfill  
{\bf  IFJPAN-IV-2021-20, December 2021}





\end{titlepage}
\newpage
\section{Introduction}
\label{intro}

The {\tt tauola} package
\cite{Jadach:1990mz,Jezabek:1991qp,Jadach:1993hs,Golonka:2003xt} for simulation
of $\tau$-lepton decays and
{\tt photos} \cite{Barberio:1990ms,Barberio:1994qi,Golonka:2005pn} for simulation of QED radiative corrections
in decays, are computing
projects with a rather long history. Written and maintained by
well-defined principal authors, they nonetheless migrated into a wide range
of applications where they became essential ingredients of
complicated simulation chains. In the following, we shall use
version of the programs which are prepared for installation
in {\tt basf2} software of the Belle II experiment. The following programs are installed in the system:
  (i) {\tt KKMC} for the $\tau$ lepton production process
  $e^-e^+ \to \tau^-\tau^+ n\gamma$
  (ii)  {\tt tauola} for $\tau$ lepton decays,
  (iii) {\tt photos} for bremsstrahlung in decays of particles and resonances,
  (iv)  {\tt photospp}, the C++ version of {\tt photos}, which at present
  is used only for supplementing events with lepton pairs, produced through
  virtual careers of the electroweak interaction in the Standard Model or through New Physics processes.

  Technical changes are not addressed, they can be found in 
  Ref.~\cite{Banerjee:2021rtn}.
Recent extensions  of phase space generators, enable not only
full implementation of bremsstrahlung-like processes where virtual photon decay
into pair of leptons, but also the possibilities of emitting dark photon or dark scalars,
now introduced into {\tt tauola} and {\tt photospp}, and instrumental for the
present talk.

Example numerical results of dark scalar implementation
into {\tt KKMC} \cite{Jadach:1999vf} $e^-e^+ \to \tau^-\tau^+ n\gamma$ event samples are presented.
From technical point of view, first the introduction of
phase space presamplers for lepton pairs originating from virtual photon
or narrow exotic resonances, either vector or scalar in nature was necessary.
Both {\tt tauola} and {\tt photos} rely
on exact phase space parametrisations. Of course results of simulations
also depend on parametrization of matrix elements, but if presamplers of phase-space are not appropriate,
efficiency of generation is poor and in extreme cases the distributions may be unreliable.

Pair emission can be also generated with other generators. That was used
for test already long time ago. Automated comparison package {\tt MC-TESTER}
\cite{Davidson:2008ma} was used to construct histograms and compare results from {\tt KORALW}
\cite{koralw:1998} with those from {\tt photos}  in tandem with {\tt KKMC}.
Such  tests are very helpful.  Now events simulated with  {\tt photos} 
were compared with events simulated with {\tt MadGraph}  for the $e^-e^+ \to \tau^- \tau^+ X\; (X \to  l \bar{l})$ process,
where X is an exotic particle motivated by new physics models,
and can be dark photon 
or dark scalar \cite{Batell:2016ove}.
The spin state of $\tau$ flips when X is scalar,
which needs to be taken into account in proper simulation of $\tau-\tau$ spin correlations.
For these developments,  distributions prepared with {\tt MC-TESTER}
were obtained, and used in the development of {\tt tauola} and {\tt photospp}.

\begin{figure}[!h]
  \begin{center}
    \includegraphics[width=0.49\textwidth]{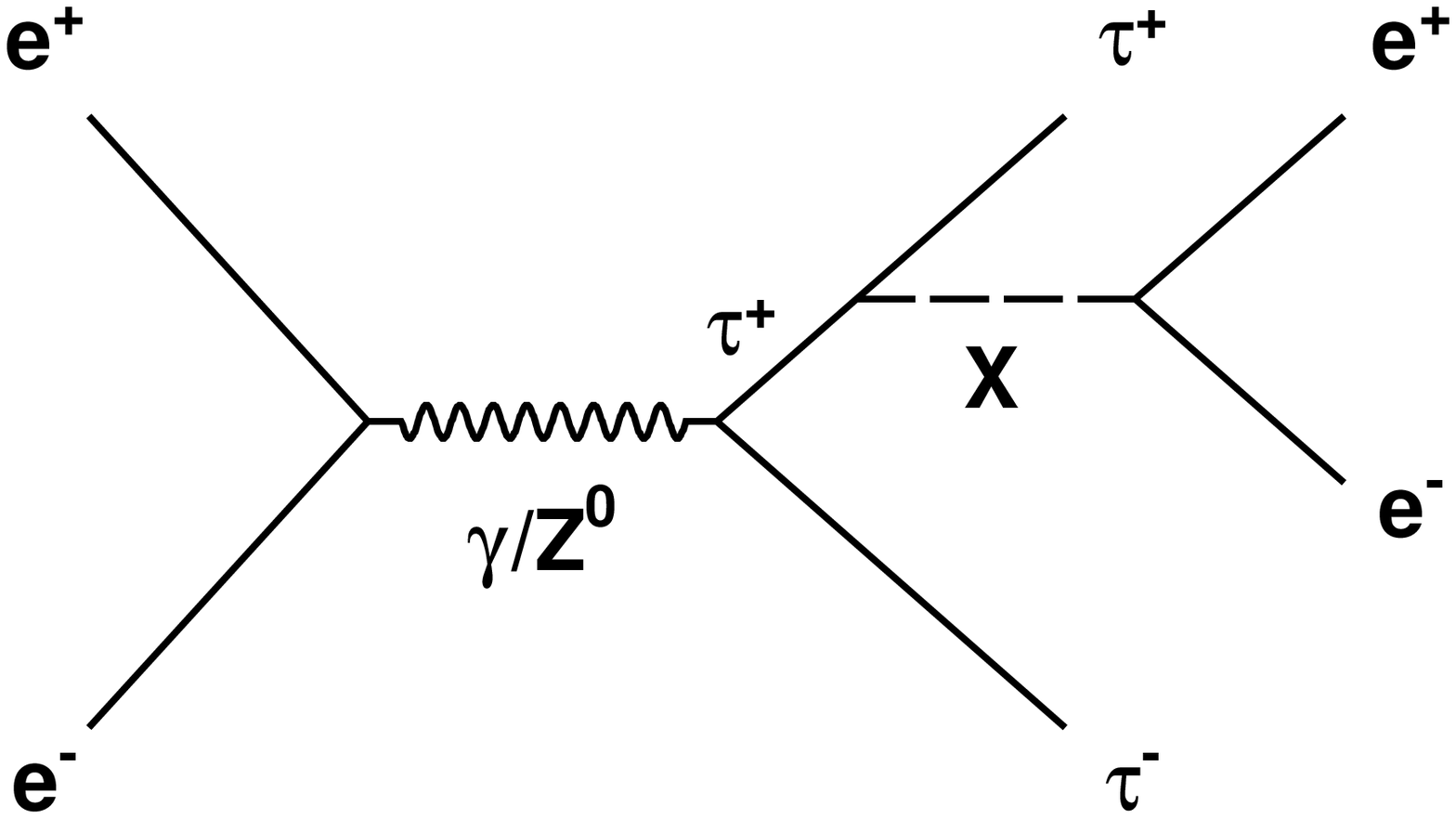}  
    \includegraphics[width=0.49\textwidth]{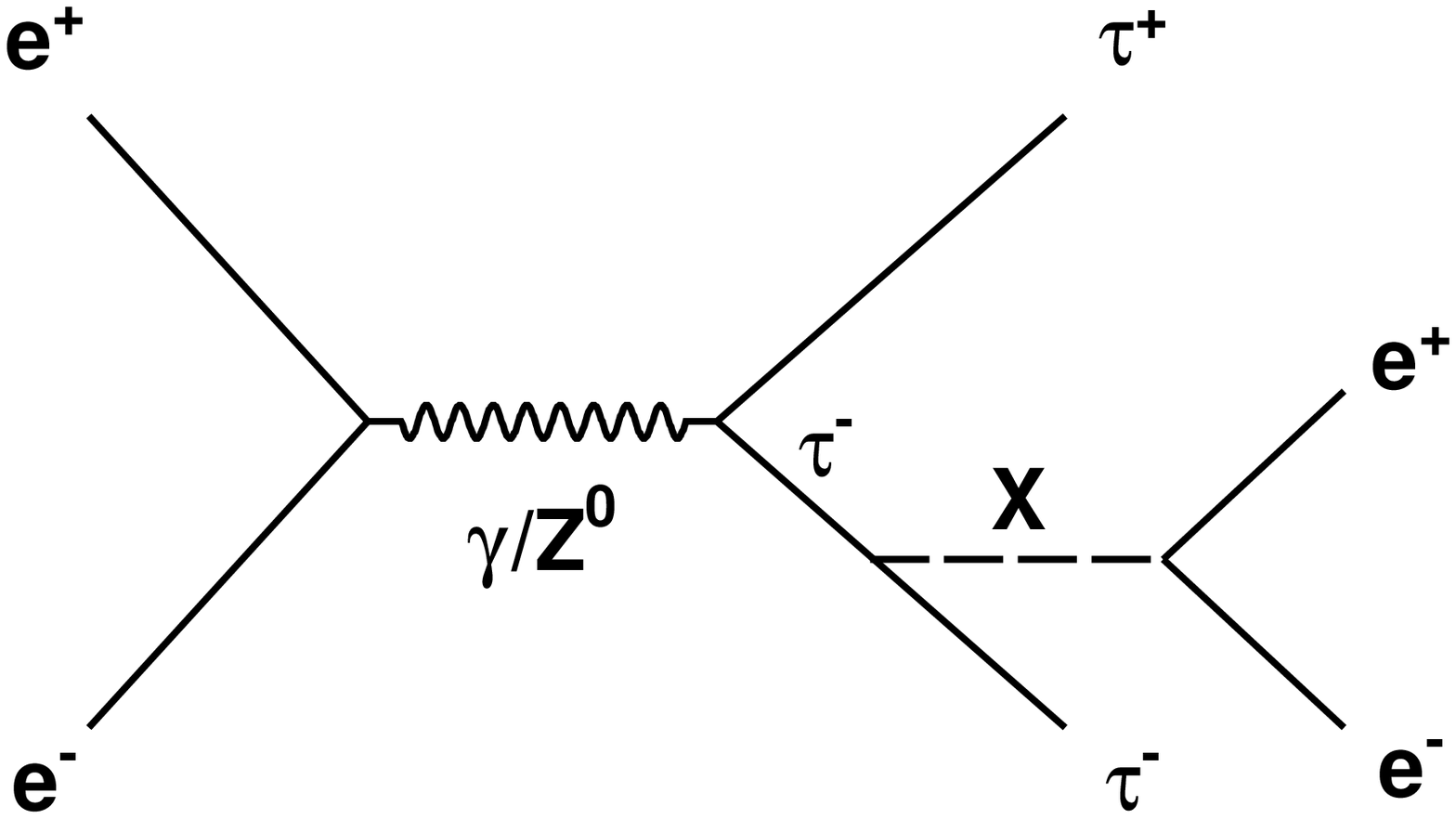}  
    \caption{Feynman diagrams for $e^-e^+ \to \tau^- \tau^+ X (\to e^- e^+)$,
      used for preparation of {\tt photospp} emission kernel.}
\label{fig:Feyn}
\end{center}
\end{figure}

Numerical tests for Standard Model pair emission algorithm are  published \cite{Antropov:2017bed}, 
and following updates on the program presented in \cite{Davidson:2010ew}.
This update opened up new possibilities, in particular,
generation of lepton pairs in the process
$e^-e^+ \to \tau^-\tau^+ n\gamma$ with $\tau$ decays and implementation into final state of
an exotic particle X motivated by new physics.
There are two reasons, why matrix elements of refs  \cite{Shuve:2014doa,Altmannshofer:2016jzy,Batell:2016ove} could not be
used directly. First is to preserve modularity of {\tt photos} Monte Carlo
design. Second is because the matrix element form must enable its interpolation for use when additional bremsstrahlung photons are present. That is why
a factorized form, with the emission factor similar to the eikonal one, was necessary to be devised and checked for the process shown in Fig.~\ref{fig:Feyn}.

  For configurations where lepton pairs from decay of exotic scalars
  or vector particles are present, approximate matrix elements were derived,
  following educated guesses.
  The approximations were then validated with {\tt MadGraph} simulation~\cite{Alwall:2014hca}
 samples.
  The best of  several variants was chosen.
  This opened up the gateway for  simultaneous inclusion of large QED effects, e.g. ISR as implemented in {\tt KKMC}. ISR effects 
  were incorporated in {\tt MadGraph} simulation using the recipe from \cite{Li:2018qnh}.
  
\begin{figure}[!h]
  \begin{center}
{ \resizebox*{0.4\textwidth}{0.22\textheight}{\includegraphics{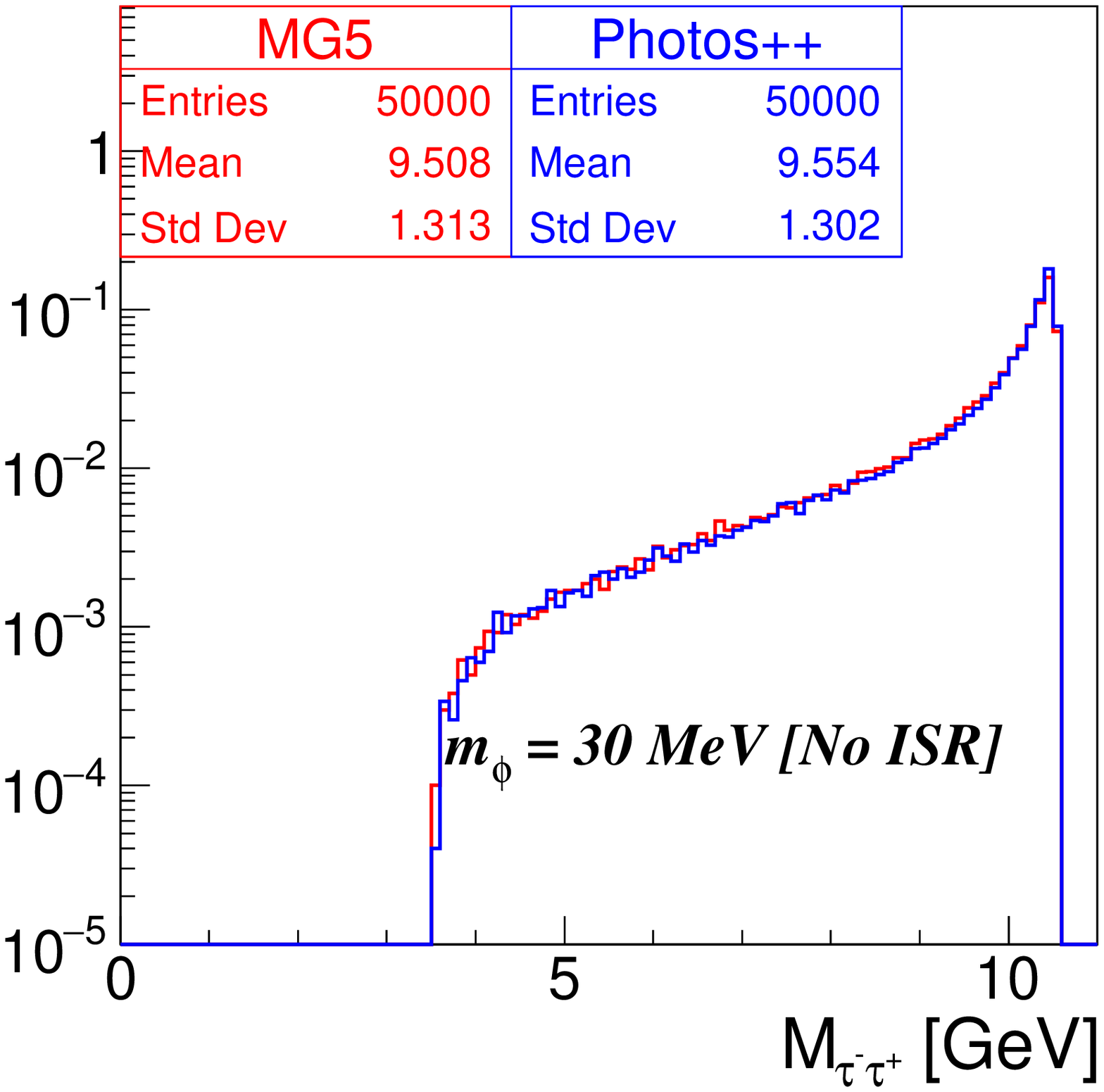}} }
{ \resizebox*{0.4\textwidth}{0.22\textheight}{\includegraphics{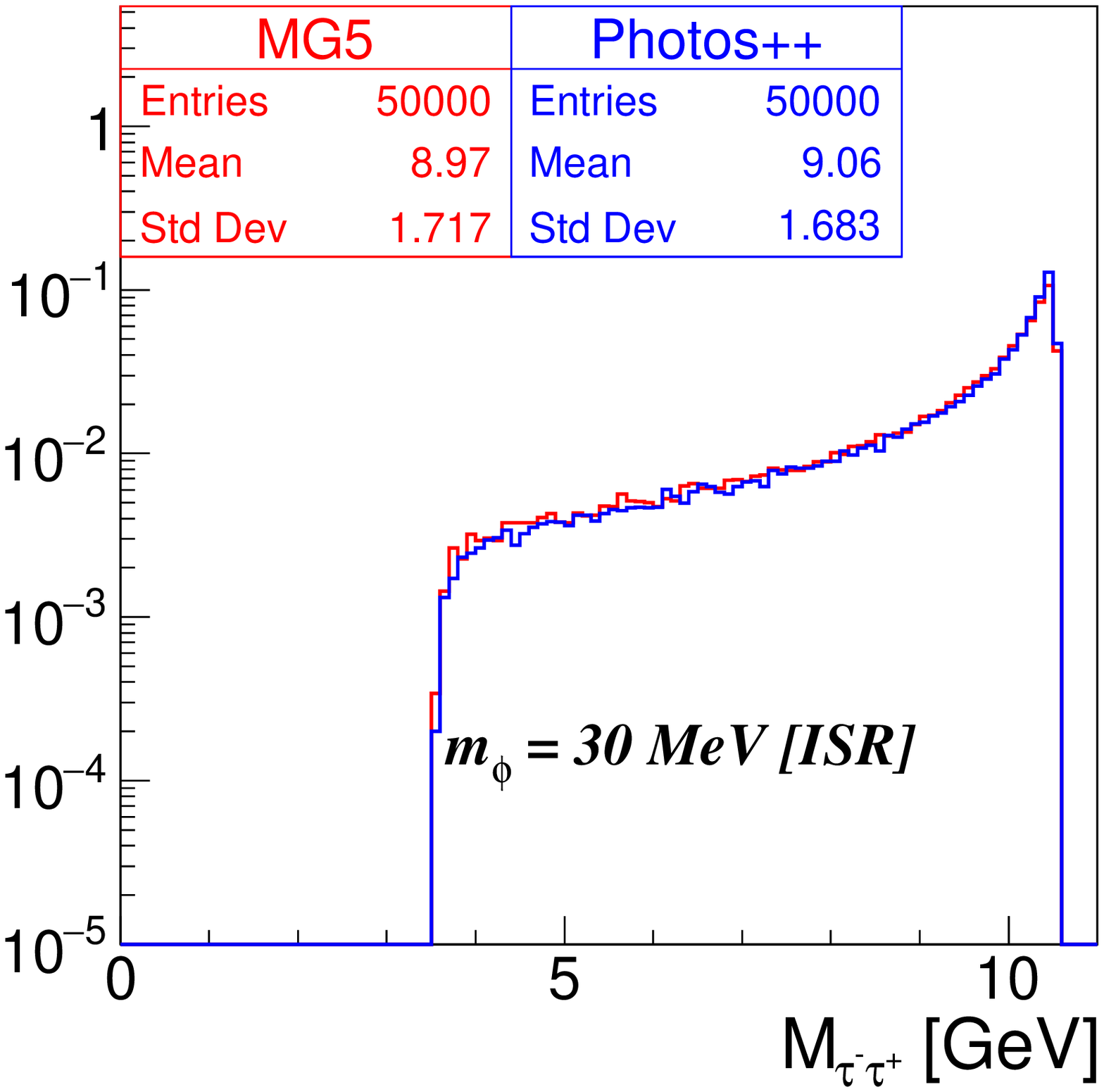}} }
{ \resizebox*{0.4\textwidth}{0.22\textheight}{\includegraphics{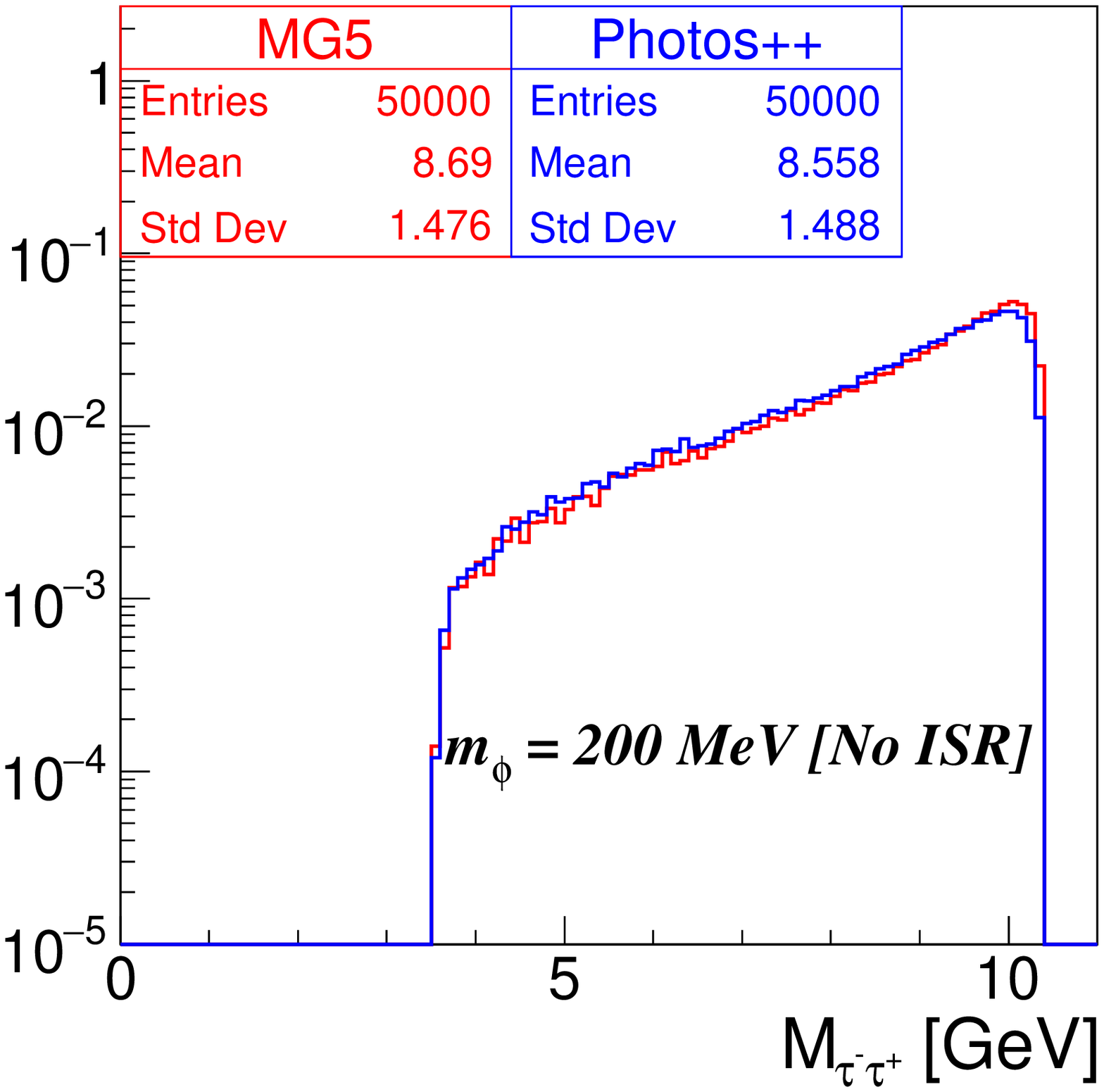}} }
{ \resizebox*{0.4\textwidth}{0.22\textheight}{\includegraphics{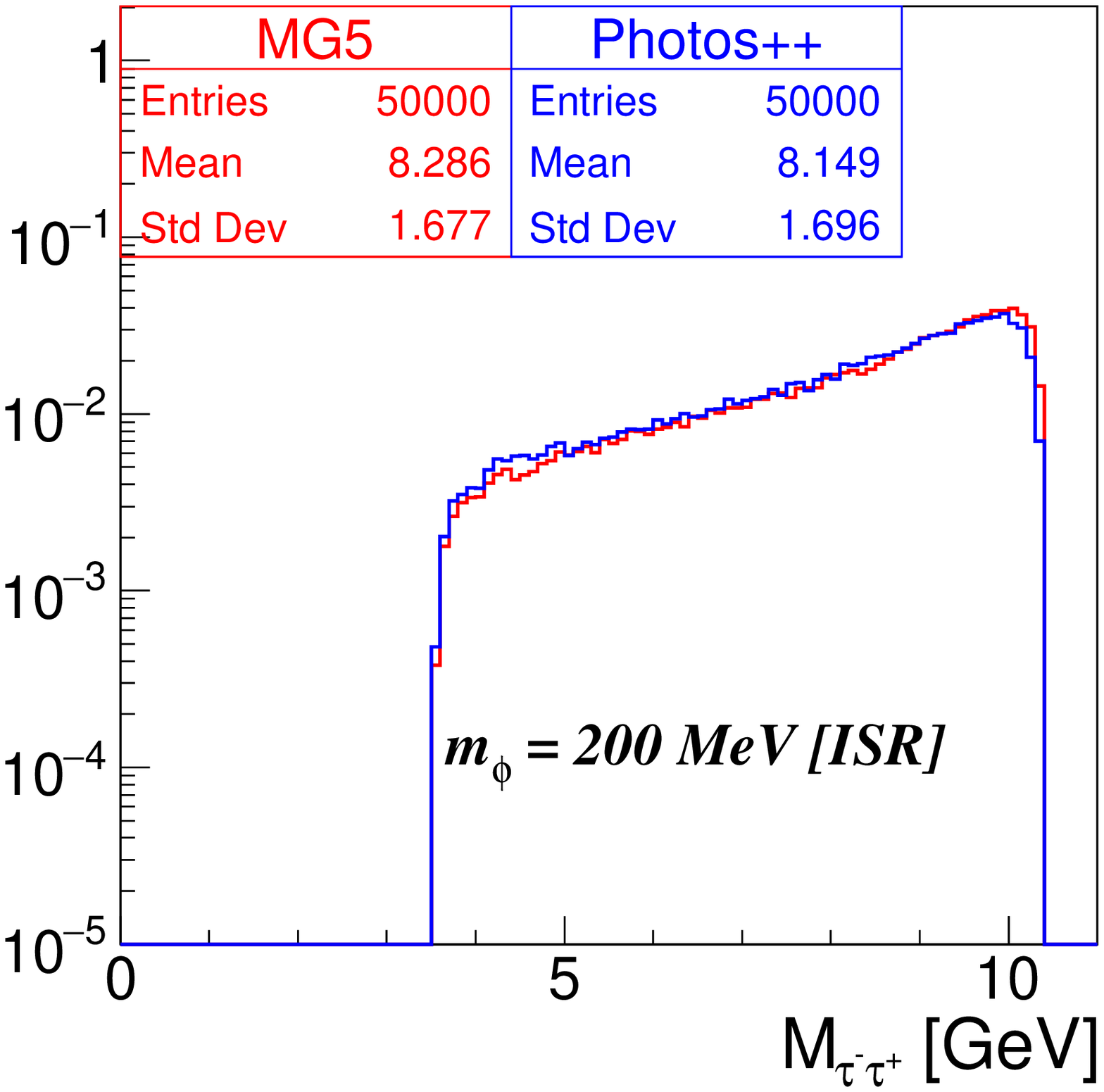}} }
\caption{$e^-e^+ \to \tau^- \tau^+ \phi_{\rm{Dark~Scalar}} (\to e^-e^+)$}\label{fig:ee}
  \end{center}
 \end{figure}

\begin{figure}[!h]
  \begin{center}
{ \resizebox*{0.4\textwidth}{0.22\textheight}{\includegraphics{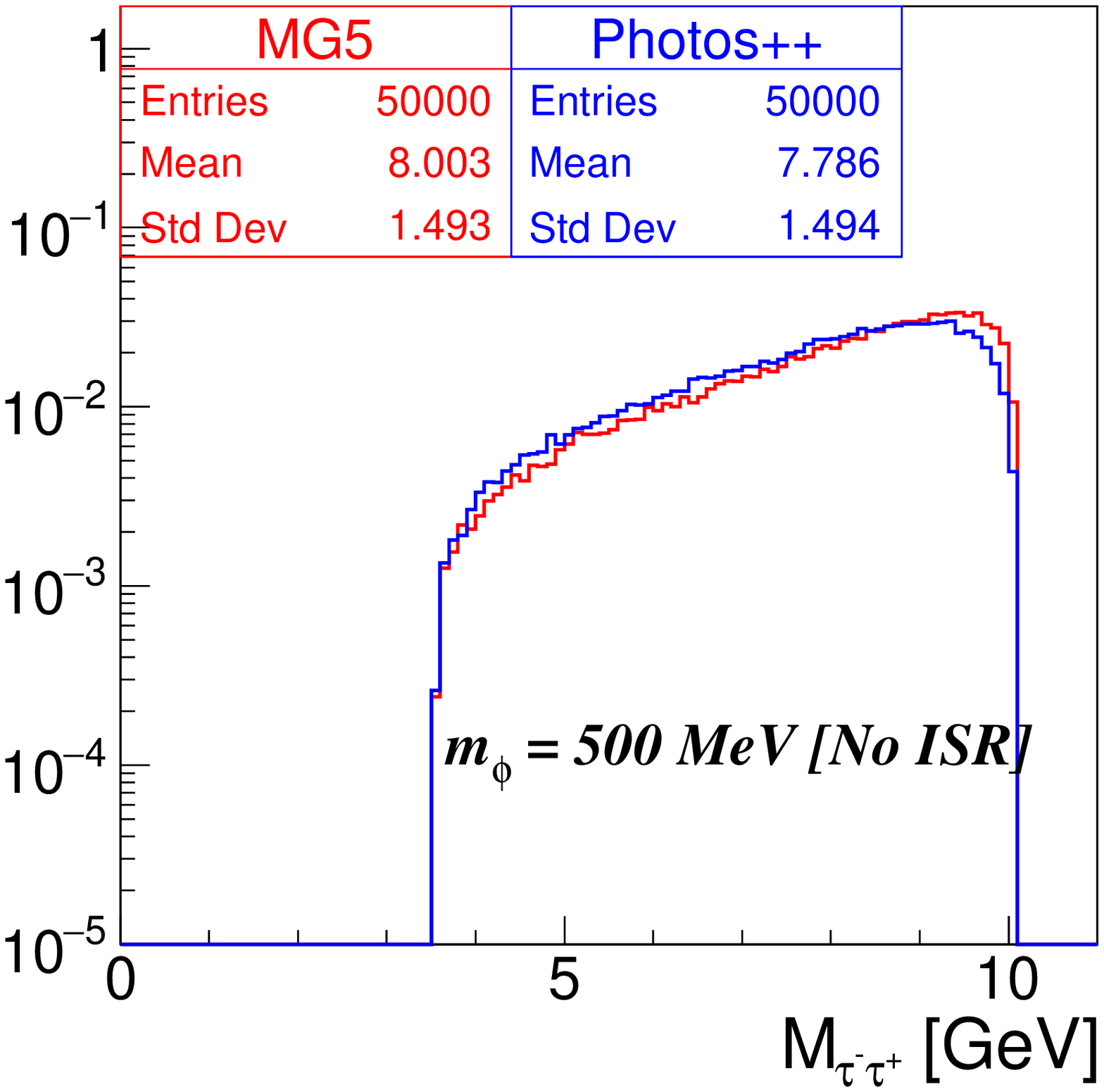}} }
{ \resizebox*{0.4\textwidth}{0.22\textheight}{\includegraphics{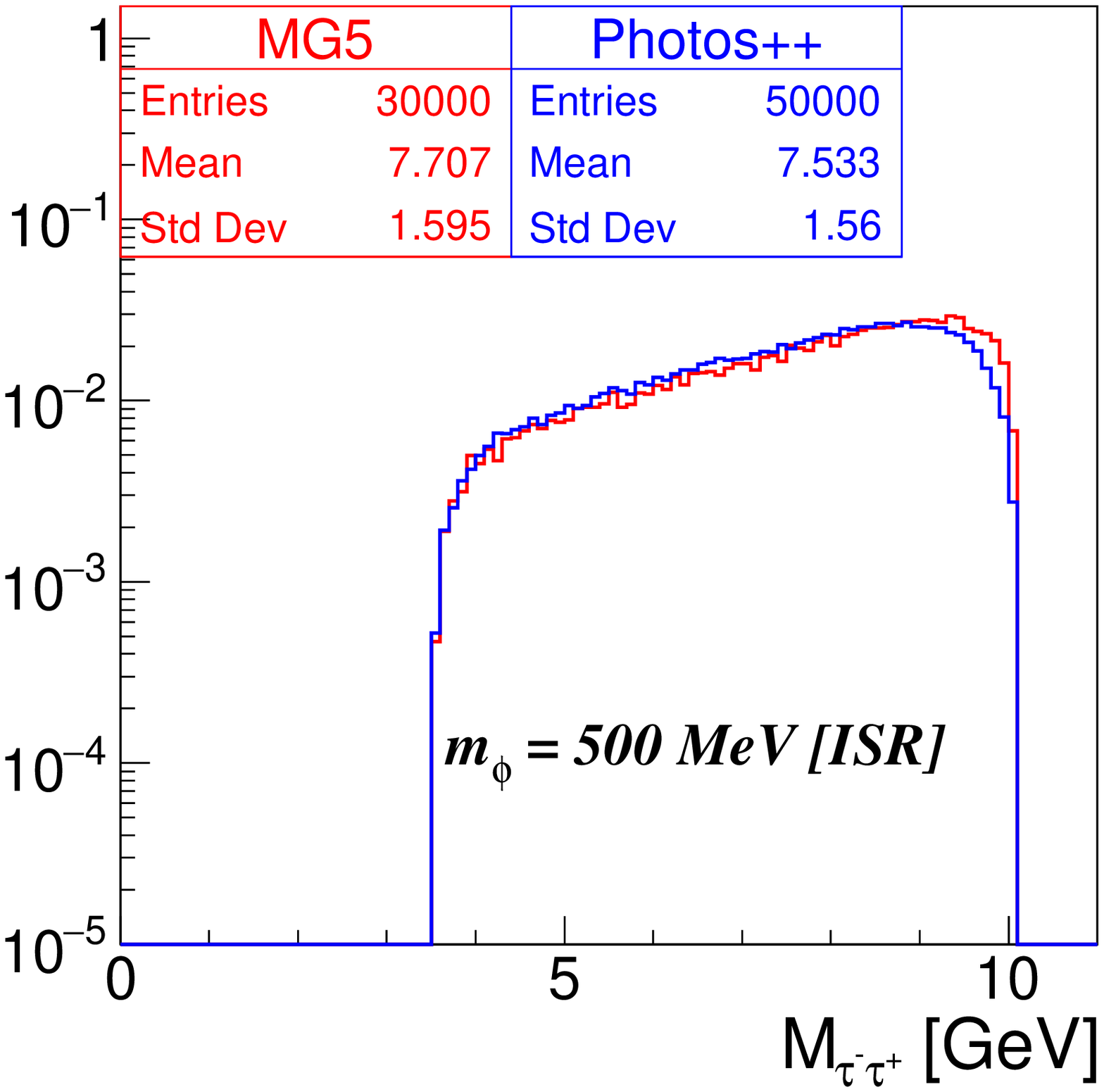}} }
{ \resizebox*{0.4\textwidth}{0.22\textheight}{\includegraphics{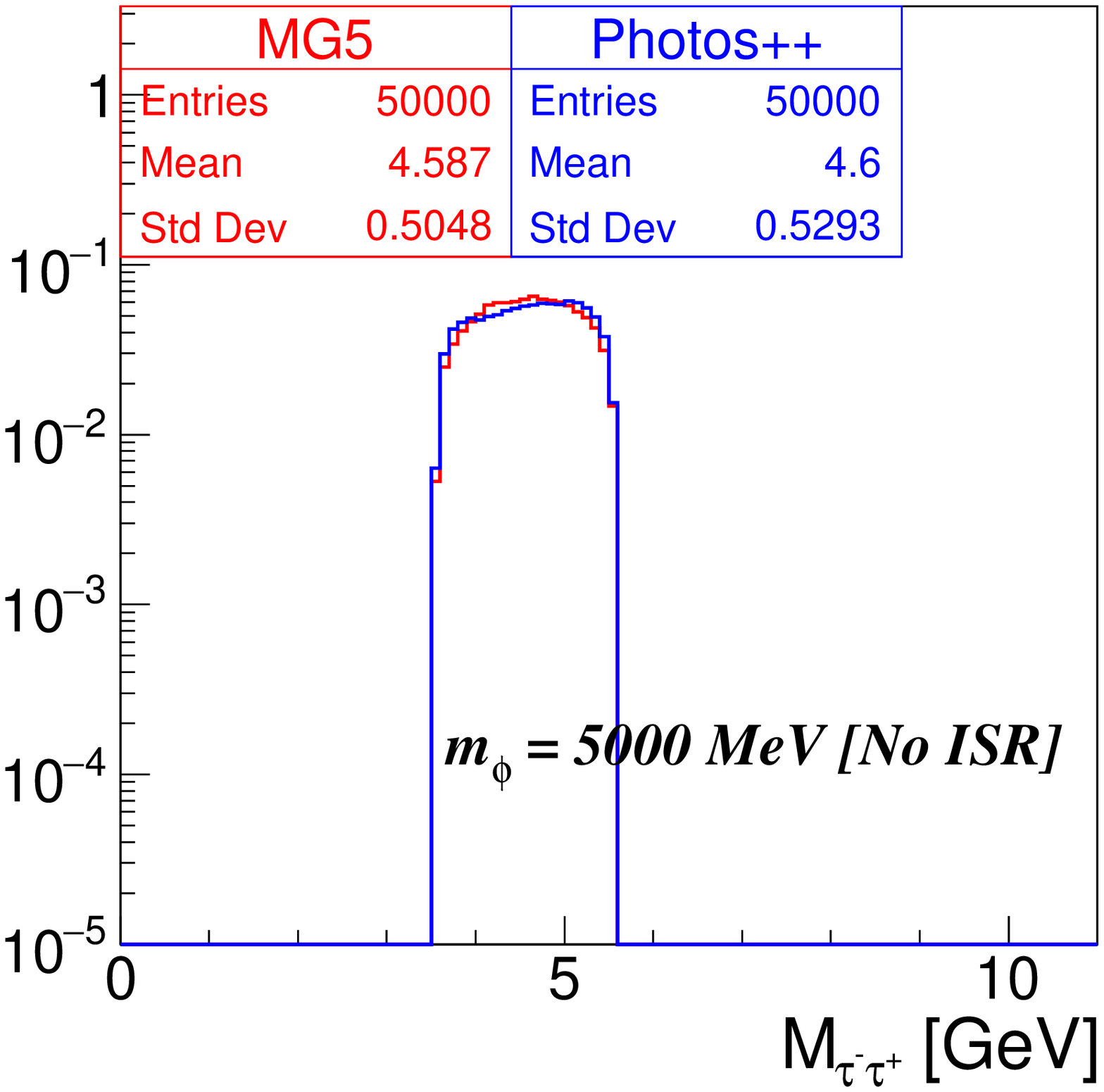}} }
{ \resizebox*{0.4\textwidth}{0.22\textheight}{\includegraphics{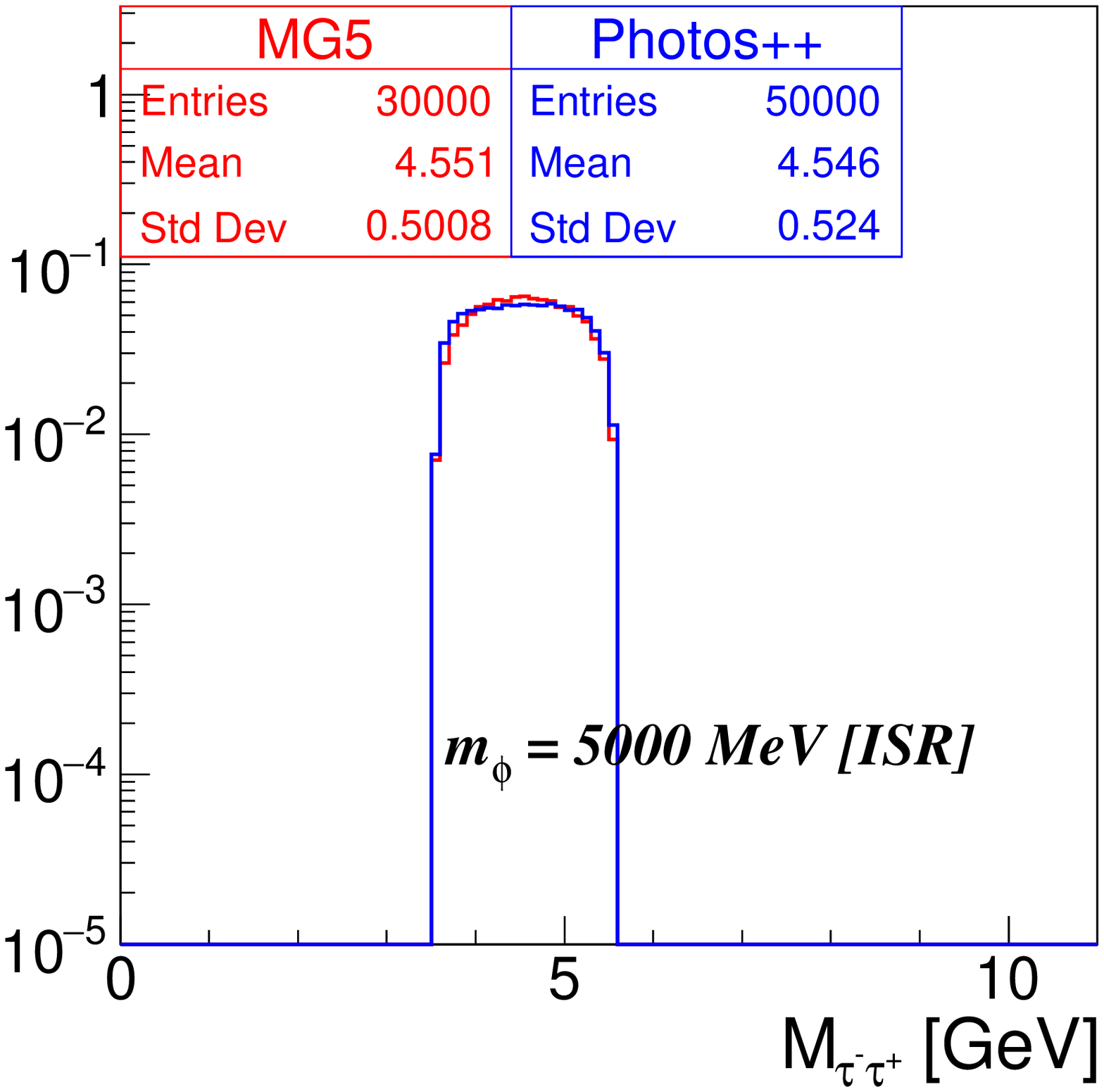}} }
\caption{ $e^-e^+ \to \tau^- \tau^+ \phi_{\rm{Dark~Scalar}} (\to \mu^-\mu^+)$ } \label{fig:mumu}
  \end{center}
\end{figure}

In fact, not only test with {\tt MadGraph} simulation~\cite{Alwall:2014hca} samples were necessary, but also several iteration of {\tt photos}
matrix elements were performed  to achieve better a simulation  tool.
Validations and choices were performed  with the help of {\tt MC-TESTER}
Shape Difference Parameter.  Final validation of  {\tt photos} was the check
on the distributions
of the recoil mass of the $\tau$-pair system
for the process $e^-e^+ \to \tau^- \tau^+ \phi_{\rm{Dark~Scalar}}$,
where the $\phi_{\rm{Dark~Scalar}}$ decays into a pair of oppositely charged electrons or muons,
as shown in Figs.~\ref{fig:ee} and~\ref{fig:mumu}.

\begin{figure}[!h]
  \begin{center}
    \includegraphics[width=0.49\textwidth]{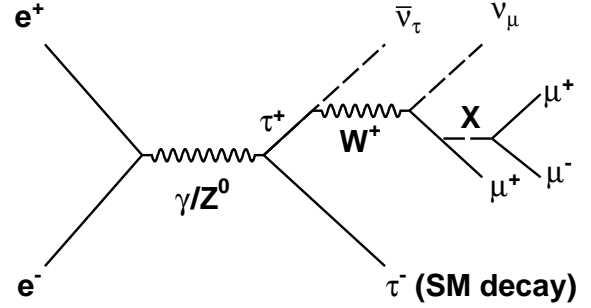}
   \caption{ Feynman diagram for $e^-e^+ \to \tau^-~({SM~decay})~ \tau^+ (\to \bar{\nu}_\tau \nu_\mu \mu^+ X (\to \mu^- \mu^+))$ process with  X emitted in $\tau^+$ decay, used in {\tt tauola}. }
   \label{fig:decay}
\end{center}
\end{figure}

  New decay modes with SM photons or Dark photons decaying to lepton pair with mass $\in$ $[50, 1500]$ MeV
  with matrix elements cross-validated with {\tt MadGraph}~\cite{Alwall:2014hca} was introduced into {\tt tauola}. In particular
  $\tau^- \to \nu_\tau \bar{\nu}_{\ell} \ell^- \ell^+ \ell^-$ decays.
    A typical Feynman diagram for such process is shown
      in Fig.~\ref{fig:decay}.

Example numerical results are given in Fig. \ref{fig:ee} and \ref{fig:mumu}, results of our simulation is compared with the one of {\tt MadGraph}.
\section{  Summary }

Final states of $\tau$ lepton pairs with bremsstrahlung photons
and dark scalar/photon (decaying to the lepton pair) were 
introduced for  study of $e^-e^+$ collisions. The $\tau$ pair production and
decay were introduced with {\tt KKMC} and  {\tt tauola}, dark scalar/photon
with {\tt photospp}. They may
correspond to virtual corrections in the calculation of anomalous g-2.

Motivation of the talk was to present solutions that may be of some use to
the broader community than just the Belle II users of {\tt KKMC}.
In principle solution
is straightforward to extend to any New Physics states which decay to a pair of
$\tau$ leptons. However alternative solution based on similar  modifications
to the {\tt KORALW} generator may be more appropriate. This is because 
contributions from  terms proportional to $\tau$ mass play a more important role
than initial state bremsstrahlung in such cases. 

\vskip 2 mm
\centerline{\bf Acknowledgments}

{\small 
This project was supported in part from funds of Polish National
Science Centre under decisions DEC-2017/27/B/ST2/01391.
}


\bibliographystyle{elsarticle-num}
\bibliography{Tauola_interface_design.bib}

\end{document}